\documentclass{appolb}%
\usepackage{graphicx}
\usepackage{amssymb}
\usepackage{amsmath}
\usepackage{bm}  
\usepackage{amsfonts}

\begin{document}
\renewcommand*{\thefootnote}{\fnsymbol{footnote}}

\title{Heavy baryons in
the Chiral Quark-Soliton Model\thanks{Presented 
at 
{\em Excited QCD}, Giardini-Naxos, Italy, October 24 -- 28, 2022.}}
\author{Micha{\l} Prasza{\l}owicz
\address{Institute of Theoretical Physics, Faculty of Physics, Astronomy and Applied Computer Science,  Jagiellonian University, \\
ul. S. {\L}ojasiewicza 11, 30-348 Krak{\'o}w, Poland.}}

\maketitle

\begin{abstract}
We review applications of the Chiral Quark Soliton Model to heavy baryons and to doubly heavy tetraquarks.
\end{abstract}


\section{Introduction \label{sec:intro}}
Chiral Quark Soliton Model ($\chi$QSM) was initially designed to describe spectra of light baryons \cite{Diakonov:1987ty}.
Nevertheless, a possibility to apply it to heavy baryons (with one heavy quark) was already
discussed by D.I. Diakonov in 2010 \cite{Diakonov:2010tf}. Six years later the present author with
collaborators \cite{Yang:2016qdz} revived the idea of Ref.~\cite{Diakonov:2010tf} extending it
to putative exotic states \cite{Kim:2017jpx,Kim:2017khv,Praszalowicz:2022hcp}, negative parity excited
baryons \cite{Polyakov:2022eub}, and even to doubly heavy tetraquarks \cite{Praszalowicz:2019lje,Praszalowicz:2022sqx}.
The aim of the present paper is to briefly summarize  these developments.

The $\chi$QSM is based  on Witten's argument ~\cite{Witten:1979kh} that in the limit $N_{\rm val}=N_{\rm c} \rightarrow\infty$
relativistic valence quarks generate
chiral mean fields represented by a distortion of  the Dirac sea, which in turn
interacts with the valence quarks,
which in turn modify the sea until a stable configuration is reached. This
configuration is called  a \emph{Chiral Quark Soliton} ($\chi$QS).  $\chi$QS
corresponds to the solution of the
Dirac equation for the constituent quarks, with fully occupied Dirac sea. Therefore
at this stage the model is fully relativistic.

For large $N_{\rm c}$ the $\chi$QS is heavy, and the quantization of the zero modes of the 
underlying {\em hedgehog} symmetry leads to the correct light baryon spectrum 
(see {\em e.g.} \cite{Petrov:2016vvl}). The effective collective Hamiltonian represents
a non-relativistic rigid rotation of $\chi$QS in the SU(3) group space, and is characterized by the soliton mass
$M_{\rm sol}$, two moments inertia $I_{1,2}$ and parameters describing chiral symmetry breaking
due to the nonzero strange quark mass.  The collective baryon wave function is given
in terms of Wigner matrix $D_{B,S}^{({\cal{ R}})*}$ where $B=(Y,T,T_3)$ corresponds to the SU(3) quantum numbers of
the baryon in representation $\cal{R}$, and $S=(Y',T',T'_3)$ is related to the soliton spin $J$. Here $Y'=N_{\rm val}/3$ selects
the allowed SU(3) representations. For light ground state baryons $\cal{R}={\bf 8}$ or {\bf 10} with  $T'=J$ and $T'_3=-J_3$,
where $T'=1/2$ for octet and 3/2 for decuplet.

\begin{figure}[h]
\centering
\includegraphics[width=10cm]{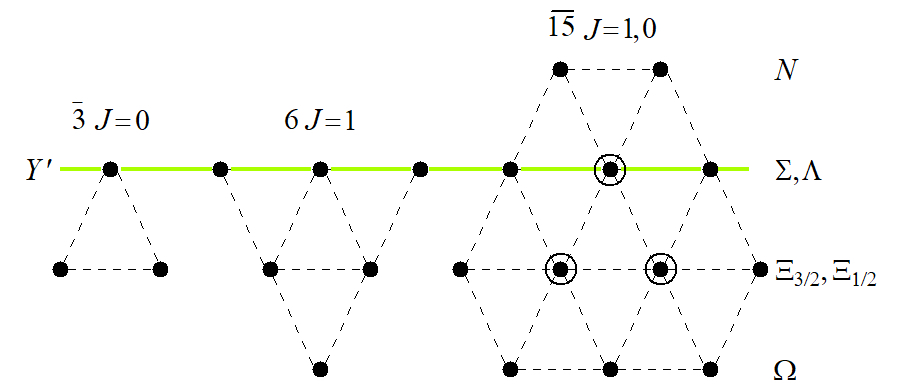} 
\vspace{-0.2cm}%
\caption{Rotational band of the $\chi$QS with one valence quark stripped off.
Soliton spin is related to the isospin $T^{\prime}$ of states on the
quantization line $Y^{\prime}=2/3$ (green thick horizontal line). 
On the right hand side we display particle names used
in the present paper. Figure from Ref.~\cite{Praszalowicz:2022hcp}.}
\label{fig:reps}%
\end{figure}

It has been initially observed in \cite{Diakonov:2010tf} that removing one valence quark leading to $N_{\rm val}=N_{\rm c}-1$
hardly changes chiral mean fields in the limit  $N_c \rightarrow \infty$, however, the quantization rule related to 
$S$ selects  $\cal{R}=\overline{{\bf 3}}$ with $J=0$
 or {\bf 6} with $J=1$, see Fig.~\ref{fig:reps}. 
 Assuming that heavy baryon consists from such a soliton and a heavy quark, one reproduces the quark
 model SU(3) pattern of ground state heavy baryons. In what follows we shall explore the resulting phenomenology.
 
\section{Positive parity heavy baryons}
States in the multiplets of Fig.~\ref{fig:reps} are degenerate in the SU(3) symmetry limit. The collective Hamiltonian has to be supplemented
 by the perturbation:
 \begin{equation}
H_{\mathrm{{sb}}}=\alpha\,D_{88}^{(8)}+\beta\,\hat{Y}+\frac{\gamma}{\sqrt{3}%
}\sum_{i=1}^{3}D_{8i}^{(8)}\,\hat{J}_{i}, \label{eq:Hsb}%
\end{equation}
where $\alpha$, $\beta$, and $\gamma$ are proportional to $m_s-m_{u,d}$
 and are
given in terms of the moments of
inertia and the pion-nucleon sigma term, see Ref.~\cite{Yang:2016qdz} for their explicit form.

Since we know the collective wave functions, it is rather straightforward to compute the mass splittings in the first order of the
perturbative expansion. The result reads
\begin{align}
M_{\overline{\mathbf{{3}}},J=0}^{Q}  &  =m_{Q}+M_{\mathrm{{sol}}}+\frac{1}{2I_{2}} +\delta_{\overline{\mathbf{{3}}}} Y,
\nonumber\\
M_{\mathbf{{6}},J=1}^{Q}  &  =M_{\overline{\mathbf{{3}}}}^{Q}+\frac{1}{I_{1}}+\delta_{\mathbf{{6}}} Y,
\label{eq:mass3bar6}
\end{align}
where $\delta_{\overline{\mathbf{{3}}}}$ and $\delta_{\mathbf{{6}}}$ are known functions of $\alpha$, $\beta$ and $\gamma$ \cite{Yang:2016qdz}.

Since the soliton in ${\bf 6}$ is quantized as spin $J=1$, we have to add spin-spin interaction between  the heavy quark and the soliton,
leading to the hyperfine splitting
\begin{align}
\delta_{\rm 6}^{({\rm h.f.})}=&
\frac{\varkappa}{m_{Q}}\left\{
\begin{array}
[c]{ccc}%
-2/3 & \text{for} & s=1/2\\
\, & \, & \,\\
+1/3 & \text{for} & s=3/2
\end{array}
\right.
\label{eq:hf6}%
\end{align}
where $s$ stands for heavy baryon spin and $\varkappa$ is a new free parameter.

Formulae (\ref{eq:mass3bar6}) and (\ref{eq:hf6}) imply Gell-Mann--Okubo equal spacing mass relations
and one relation allowing to compute $\Omega_Q^*(s=3/2)$ mass:
\begin{align}
M_{\Omega_{Q}^{\ast}} 
= 
 2M_{\Xi^{\prime}_{Q}}
 +M_{\Sigma_{Q}^{\ast}} 
 -2 M_{\Sigma_{Q}}.
\label{eq:OMpred}
\end{align}
Equation~(\ref{eq:OMpred}) yields $(2764.5\pm3.1)$~MeV for
$M_{\Omega_{c}^{\ast}}$, which is 1.4~MeV below the experiment, 
and predicts~\cite{Yang:2016qdz}
\begin{align}
M_{\Omega_{b}^{\ast}}=6076.8\pm2.25\;\mathrm{MeV}.
\label{eq:Omstb2}
 \end{align}

In order to compute masses in a model independent way (in a sense of \cite{Adkins:1984cf}) one can try
to extract the parameters from the light baryon spectra \cite{Yang:2016qdz} or entirely from the heavy quark
sector \cite{Praszalowicz:2022hcp}. In both cases the description of the data is very good, although in the first case
some modification of the parameters proportional to $N_{\rm val}$ is required  \cite{Yang:2016qdz} in rough
agreement with model calculations \cite{Kim:2018xlc,Kim:2019rcx}.

The model allows to compute decay widths 
$B_1 \rightarrow B_2 + \varphi$ in no recoil approximation \cite{Kim:2017khv}. The decay operator can be computed
via the Goldberger-Treiman relation
\begin{align}
\mathcal{O}_{\varphi}&=\frac{1}{2F_{\varphi}}\left[  -\tilde{a}_{1}%
D_{\varphi\,i}^{(8)}-a_{2}\,d_{ibc}D_{\varphi\,b}^{(8)}\hat{J}_{c}-a_{3}%
\frac{1}{\sqrt{3}}D_{\varphi\,8}^{(8)}\hat{J}_{i}\right]  \,p_{i} .  
\label{eq:Oai}%
\end{align}
Constants $\tilde{a}_1,\, a_{2,3}$ that enter Eq.~(\ref{eq:Oai})
can
been extracted from the semileptonic decays of the baryon octet \cite{Kim:2017khv}. Here $\varphi$
stands for a pseudoscalar meson, $F_{\varphi}$ for the pertinent decay constant and
$p_i$ for meson momentum. Again the results for the decay widths are in very good
agreement with data \cite{Kim:2017khv}.

The model has been also applied to compute other quantities: magnetic moments \cite{Yang:2018uoj},
form factors \cite{Kim:2018nqf,Kim:2019wbg,Kim:2020uqo,Yang:2020klp}, isospin splittings \cite{Yang:2020klp},
nuclear matter effects on heavy baryon masses \cite{Won:2021pwb,Ghim:2022zob}
and spin content of heavy baryons \cite{Suh:2022atr}.

\section{Exotica \label{sec:exotica}}

Quantization condition $Y'=N_{\rm val}/3=2/3$ selects not only $\cal{R}=\overline{{\bf 3}}$ and
 {\bf 6} but also exotic  $\overline{{\bf 15}}$ pentaquarks, which can be quantized as $J=0$ or 1, see Fig.~\ref{fig:reps}.
 It turns out that $J=1$ multiplet is lighter~\cite{Kim:2017jpx}. Adding a heavy quark leads to two hyperfine split multiplets of
 $s=1/2$ and $3/2$. It has been proposed in Ref.~\cite{Kim:2017jpx} that two narrowest $\Omega_c$
 baryons discovered by the LHCb Collaboration in 2017 \cite{LHCb:2017uwr}, 
 namely $\Omega_{c}^{0}(3050)$ and $\Omega_{c}^{0}(3119)$ belong to $\overline{{\bf 15}}_{J=1}$.
This
assignment has been motivated by the fact that their
hyperfine splitting is equal to the one of the ground state sextet, which is in the same rotational band,
and has been further reinforced by
the calculation of their widths~\cite{Kim:2017khv}, which vanish for $N_{\rm c} \rightarrow \infty$ \cite{Praszalowicz:2018upb}.

Introducing new exotic multiplets, in itself very attractive, is nevertheless a phenomenological challange,
as we have to explain why 43 new exotic states have not been so far observed. Recently in Ref.~\cite{Praszalowicz:2022hcp}
we have shown that states in $\overline{{\bf 15}}_{J=1}$ are in fact very narrow, and on the contrary $\overline{{\bf 15}}_{J=0}$
is very broad.
The identification of exotica requires
 dedicated experiments. Multipurpose searches can easily miss narrow or wide exotic states. Interestingly the lightest 
 nucleon-like pentaquark in $\overline{{\bf 15}}_{J=0}$ (see Fig.~\ref{fig:reps}) decays only to
 to the $N_c$ state in $\overline{\boldsymbol{15}}_{J=1}$, which is semi-stable.

 \section{Negative parity heavy baryons}
 
 In Sec.~\ref{sec:exotica} we have argued that two out of five $\Omega_c$ states discovered by the LHCb can be interpreted as pentaquarks. 
 In Refs.~\cite{Kim:2017jpx,Polyakov:2022eub} the remaining three have been interpreted as members of negative parity sextets.
 
 Negative parity baryons appear in the $\chi$QSM as rotational bands of the excited soliton \cite{Diakonov:2010tf,Diakonov:2013qta},
 and are therefore analogous to the diquark excitations (so-called $\rho$ modes) in the
quark language. Diquark-heavy~quark excitations, referred to as  $\lambda$ modes, are in the present approach supperssed in the large
$N_c$ limit \cite{Polyakov:2022eub}.

 In the excited $\chi$QS the empty valence level in the light sector can be taken by a quark excitation from one of the filled sea levels.
 If such a state has negative parity, the soliton itself is parity odd. The rotational band corresponds to the same SU(3) representations
 as the ground state, however the soliton spin is no longer equal to $T'$. The situation is more complicated \cite{Diakonov:2013qta}. Namely,
 the soliton spin $J$ coupled to $T'$ has to be equal to the {\em grand spin} $K=T+S$. Here $T$ and $S$ denote the isospin and the spin of
 the quark level in question. This condition follows from the {\em hedgehog} symmetry. For the ground state $K=0$, and we recover quantization
 from Sec.~\ref{sec:intro}.
 
We assume that the first sea level has $K^P=1^-$. The SU(3)  $\overline{{\bf3}}$ has  $T'=0$, and therefore negative parity soliton has $J=1$.
Adding a heavy quark we get two hyperfine split antitriplets with spin 1/2 and 3/2. This pattern is clearly seen in data \cite{Polyakov:2022eub,PDG}.

The sextet has $T'=1$ hence $J=0,1$ and $2$. Therefore heavy  negative parity baryons come in three spin submultiplets: ($J=0,s=1/2$),
($J=1,s=1/2,3/2$) and ($J=2,s=3/2,5/2$). In Ref.~\cite{Kim:2017jpx} the remaining three LHCb $\Omega_c^0$ states have been interpreted
as members of $J=0$ and 1 submultiplets, and it has been argued that $J=2$ states are very broad. Phenomenology of charm and baryon sextets has been in 
detail discussed in Ref.~\cite{Polyakov:2022eub}, where several scenarios of possible assignments of existing states and predictions for the 
remaining ones have been presented. Within these scenarios possible two body decay patterns have been discussed, and
the arguments have been given why some states have not been seen in the two-body mass distributions. 

\section{Doubly heavy tetraquarks}

Following Ref.~\cite{Gelman:2002wf} it has been already observed in  \cite{Praszalowicz:2019lje} that one can replace heavy quark by 
a heavy antidiquark without modifying the soliton. In this case one obtains the family of doubly heavy tetraquarks $q_1 q_2 \bar{Q}\bar{Q}$
in a color singlet, since a symmetric heavy diquark is in color triplet as is a single quark. A charm tetraquark infinitesimally
below the threshold has been recently observed by the LHCb \cite{LHCb:2021vvq,LHCb:2021auc}.
Because of Pauli principle $\bar{Q}\bar{Q}$ has spin 1.
The model admits spin 1 antitriplet
 falvor multiplet and ${\bf 6}$ of spin 0,1 and 2. Whether such a system is bound depends on the 
$\bar{Q}\bar{Q}$ dynamics, which has to be separately modelled. Assuming that the diquark mass is equal to the sum of the heavy
quark masses one gets overbinding \cite{Praszalowicz:2019lje}. Calculating diquark mass from the Schr{\"o}dinger equation with Cornell potential,
treating the Coulomb part as a perturbation,
one gets that the charm $\overline{{\bf3}}$ tetraquark is approximately 70~MeV above the $D^* D$ threshold, while the bottom one is bound.
 For details we refer the reader to Ref.~\cite{Praszalowicz:2019lje}.

  \section*{Acknowledgments}
\noindent  This work was supported by the Polish NCN grant 2017/27/B/ST2/01314.
 The author
acknowledges stimulating discussions with late Maxim V. Pol{\-}ya{\-}kov who participated
in this project for many years until his premature death on August 25, 2021.

\end{document}